\documentclass[fleqn,10pt]{wlscirep}
\usepackage[utf8]{inputenc}
\usepackage[T1]{fontenc}
\usepackage{xcolor}
\usepackage{ulem} 

\definecolor{addcolor}{rgb}{0.8, 0.0, 0.0}   
\definecolor{delcolor}{rgb}{0.0, 0.0, 0.8}   
\definecolor{modcolor}{rgb}{0.0, 0.0, 0.8}   
\definecolor{commentcolor}{rgb}{0.6, 0.4, 0.0} 

\title{On the coincidence between the close passage of HD7977 and the Pliocene-Pleistocene transition}

\author[1,2,*]{Zhuoya Cao}
\author[2]{Abraham Loeb}
\author[2]{Morgan MacLeod}
\affil[1]{Xingjian College, Tsinghua University, Beijing, 100084, China}
\affil[2]{Institute for Theory and Computation, Center for Astrophysics, Harvard \& Smithsonian, Cambridge, MA, 02138, USA}

\affil[*]{corresponding author: caozy21@mails.tsinghua.edu.cn}

\begin{abstract}

The Oort Cloud's dynamical evolution is significantly influenced by both the galactic tide and stellar flybys. This study investigates the particular case of HD7977's close encounter 2.47 Myr ago, which likely repopulated the Inner Oort Cloud and potentially triggered a significant comet shower on Earth. Our results demonstrate that the shower's intensity strongly depends on HD7977's impact parameter ($b$), with possible flyby distances ranging from 2,300 AU to $\sim$ 13,000 AU. For the closest approach ($b \sim 2,300$ AU), the terrestrial impact probability of 1 km comets increases by an order of magnitude compared to the steady state, slightly exceeding the asteroid impact probability at this size scale. We propose an analytical method to compute the probability of comet showers impacting Earth, which saves considerable computation time compared to N-body simulations. We identify a threshold diameter $D_0 = 2.25$ km for which yields $P = 1$ in our model, with $D_0$ following a logarithmic dependence on $b$. These findings suggest that HD7977's flyby may have caused an enhanced comet flux during the Pliocene-Pleistocene transition, which could plausibly be related to the environmental changes at this era.

\end{abstract}
\begin{document}

\flushbottom
\maketitle

\footnotetext{Accepted for publication in \textit{Scientific Reports}.}
%
%

\section*{Introduction}


The observed characteristics of long-period comets (LPCs) suggest that a distant reservoir of small bodies, the Oort Cloud, surrounds the solar system, extending up to $\sim 10^5$ astronomical units (AU) \cite{oort1950structure}. The "Dirty snowball" model \cite{whipple1950comet}, supported by ultraviolet observations of hydrogen halos from sublimating ices \cite{biermann1964mechanismen}, reinforces the idea that comets are icy remnants from the solar system’s formation, their nuclei stored in the Oort Cloud \cite{oort1979origin}. Observations of comet Halley demonstrate that its nucleus structure is predominantly governed by non-volatile dust \cite{keller1989comets}. Investigations of the comet 81P/Wild 2 show that it contains an abundance of silicate grains that are much larger than predictions of interstellar grain models, and many of these are high-temperature minerals that appear to have formed in the inner regions of the solar nebula \cite{brownlee2006comet}. Recent work also reports organic matter in the comet \cite{zolensky2024composition}. 

The observed inner edge of the Oort Cloud is at $\sim2\times10^{4}$ AU, beyond which it is called outer Oort Cloud (OOC). There is a more massive reservoir inside the outer Oort Cloud named inner Oort Cloud (IOC, also "Hills cloud") \cite{tremaine2023dynamics}. The IOC dynamically feeds comets into the OOC via stellar and planetary perturbations, and also the galactic tide \cite{hills1981comet} \cite{duncan1987formation}. Comets from IOC cannot always be observed in the inner solar system because when  when orbits reach a high eccentricity, comets are destroyed by sublimation or scattered away by gravitational encounters with the planets \cite{hills1981comet}.

Gravitational perturbations from passing stars and galactic tides continually inject comets from the OOC into the inner solar system, replenishing those lost to disintegration or planetary scattering \cite{duncan1987formation}. These introduce the steady-state flux of new comets into the inner solar system. Close stellar encounters might induce sudden comet showers \cite{hansen2022unbound} and the main contributor to comet shower is IOC \cite{heisler1990monte}. Comets from the IOC only enter the inner solar system in intense, short-lived showers triggered by rare close stellar encounters, and the flux of comets during a shower may be as high as 20 times higher than the steady-state rate\cite{duncan1987formation}.   

The population of the Oort Cloud is uncertain due to observational limits. But space missions designed to discover transiting exoplanets, have the potential to detect Oort Cloud objects by occultations of background stars \cite{ofek2010detectability}. Recent efforts to such missions are reported promising, with occultation data being collected \cite{nir2023search} \cite{hitchcock2024high}. Current estimates of the OOC comet population primarily rely on observations of LPCs \cite{wiegert1999evolution} and it is estimated that there are about $10^{12}$ comets with diameters lager than 1 km \cite{morbidelli2005origin}. There are few observational constraints on the IOC \cite{kaib20092006}, and the IOC population is mainly inferred from numerical simulations of solar system formation. While LPC fluxes provide empirical constraints on the OOC, the IOC’s poorly observable reservoir is mainly constrained theoretically \cite{hills1981comet}\cite{duncan1987formation}\cite{dones2004oort}\cite{kaib2008formation}\cite{vokrouhlicky2019origin}\cite{wajer2024oort}. The current ratio of comets in the IOC to the OOC remains uncertain, with conflicting estimates from various models. Hills (1981)\cite{hills1981comet} proposed estimated a mass ratio of $M_{\text{IOC}}/M_{\text{OOC}}\sim 200$. This implies two orders of magnitude more comets in the IOC than the OOC and that the mass of the IOC is a few tens of earth masses. Duncan et al. (1987) assumed that the IOC may contain up to five times more comets than the classical OOC \cite{duncan1987formation}. Dones et al. (2004) suggested a comparable population between the IOC and OOC based on simulations with no cluster environment \cite{dones2004oort}. Observations suggest most stars originate in clusters embedded in giant molecular clouds \cite{lada2003embedded}. Brasser et al. (2006) found that at a mean stellar density $\langle \rho \rangle \gtrsim 10 M_{\odot} \text{pc}^{-3}$, the median distance of the comet population from the Sun scales approximately as \cite{brasser2006embedded} $\propto \langle \rho \rangle^{-1/2}$. Kaib \& Quinn (2008) studied the formation of the Oort cloud in open cluster environments. They highlighted the dependence of the IOC/OOC ratio on the density of initial cluster where the solar system was born, showing dynamic variations in population ratios ranging from 1 to 10 for different cluster densities \cite{kaib2008formation}. Recent simulations by Vokrouhlický et al. (2019) and Wajer et al. (2024) emphasized the role of stellar perturbations and dynamical evolution, further complicating definitive estimates. Vokrouhlický et al. (2019) derived a comparable number of the IOC and OOC comets \cite{vokrouhlicky2019origin}, while Wajer et al. (2024) found an IOC/OOC ratio of 1.07, 3.02, 1.69 and 13.16 based on four models of initial cluster\cite{wajer2024oort}. Overall, the IOC/OOC comet ratio remains unresolved, with models ranging from parity to orders-of-magnitude differences. Given that the IOC is dynamically replenished by comets escaping the OOC \cite{hills1981comet}, it is reasonable to assume a higher cometary population in the IOC. In this work, an IOC/OOC population ratio of five was adopted, in accordance with Duncan’s (1987) result and accounting for replenishment mechanisms that favor IOC retention over OOC depletion. Thus, if there are about $10^{12}$ comets with diameters lager than 1 km in the OOC, there are about $5 \times 10^{12}$ comets in the IOC. 

The hypothesis that close stellar encounters perturbed the Oort cloud, triggering prolonged comet showers, provides a plausible mechanism for mass extinctions \cite{hut1987comet}. Geochemical evidence suggests that a comet shower rather than asteroid impact might have caused two large impacts in the late Eocene \cite{farley1998geochemical}. Although a later study by Tagle and Claeys (2004) suggests that at least one of those impacts was asteroidal\cite{tagle2004comet}, they have not entirely ruled out the possibility of comet showers producing Earth impacts, suggesting that further investigation of the potential effects on Earth resulting from the HD 7977-induced comet shower may be warranted. 

The influence of a stellar flyby on the Oort Cloud comets depends sensitively on its velocity, mass and distance. Dynamically important Oort Cloud perturbers may be lurking among nearby stars \cite{mamajek2015closest}. Dybczynski (2002) presented the detailed analysis of the result of single stellar passage near or through the Oort cometary cloud on different cloud models \cite{dybczynski2002simulating}. Here we focus on the star HD7977. Gaia DR3 data suggest that this $1M_\odot$ star may have passed within 2300 AU from the Sun 2.47 Myr ago \cite{dybczynski2024hd}, although uncertainties in the star’s proper motion necessitate further refinements with future Gaia data releases. Such a close stellar flyby could have perturbed the objects on less tight orbits. Dybczynski et al. (2024) examined noticeable changes in the motion of all Solar System bodies, especially long-period comets (LPCs) and transneptunian objects (TNOs) from HD7977's flyby. For LPCs, they primarily examined changes in the specific orbits of several comets, utilizing high-precision Gaia data. Their study did not extensively address large-scale statistical analyses about LPCs, nor did it systematically evaluate the probability of Earth impacts from comets. Separately, Kaib and Raymond (2024) investigated the effects of the HD 7977 encounter on Earth's orbit, suggesting this flyby could potentially reveal new sequences of Earth's past orbital evolution beyond 50 million years ago, including the Paleocene–Eocene Thermal Maximum\cite{2024ApJ...962L..28K}. 

Our study primarily examines the statistically significant global dynamical evolution of the entire Oort Cloud system following the stellar flyby, rather than focusing on the specific observed LPCs. The close flyby of HD 7977 is modeled to have substantially perturbed the IOC, potentially leading to an enhanced comet flux during the Pliocene-Pleistocene transition. This study quantifies the large-scale effects of HD 7977's flyby through changes in the overall perihelion distance distribution. An advancement over previous studies of comet showers is the use of a computationally efficient method to analytically calculate the probability of orbital crossings between Earth and comets, thereby estimating the probability of comet impacts on Earth. While the potential for stellar flybys to trigger comet showers and influence impact rates on Earth was established theoretically decades ago\cite{hut1987comet}, the study of these events has largely remained in the realm of generalized scenarios. The recent precise astrometric data from the Gaia mission have now transformed this landscape, enabling the identification of specific stellar encounters with our Solar System. Our work leverages this opportunity by focusing on the close passage of HD 7977. We thus move beyond the generic question of if such an event can occur to the specific one: given the actual mass, trajectory, and timing of HD 7977, under what conditions could it have generated a significant comet shower, and what would the subsequent implications for Earth have been? This study therefore endeavors to connect a well-known theoretical phenomenon with a concrete astronomical event to explore its specific role in our planetary history. While independent geological verification remains a necessary next step, our results are consistent with the hypothesis that a comet shower associated with the HD 7977 flyby could have been a contributing factor during the Pliocene-Pleistocene transition.

\section*{Results}


When encountered by a $1M_\odot$ close flyby of HD7977, the comets near the trajectory of the passing star are highly disturbed. The REBOUND simulation package \cite{rein2012rebound} was employed to study the resulting gravitational scattering. In our models, the IOC was set as a disk of massless bodies with semi-major axes following log-uniform distribution ranging from 2,000 AU to 5,000 AU, with inclination distribution following $\mathcal{N}( \pi/6, \pi/18)$, where $\mathcal{N}(\mu, \sigma)$ denotes a normal (Gaussian) probability distribution characterized by mean $\mu$ and standard deviation $\sigma$, and a sphere following log-uniform distribution ranging from 5,000 AU to 20,000 AU, while the OOC was a 20,000 AU to 100,000 AU log-uniform sphere \cite{vokrouhlicky2019origin}. 

\begin{figure}[ht]
\centering
\includegraphics[width=\linewidth]{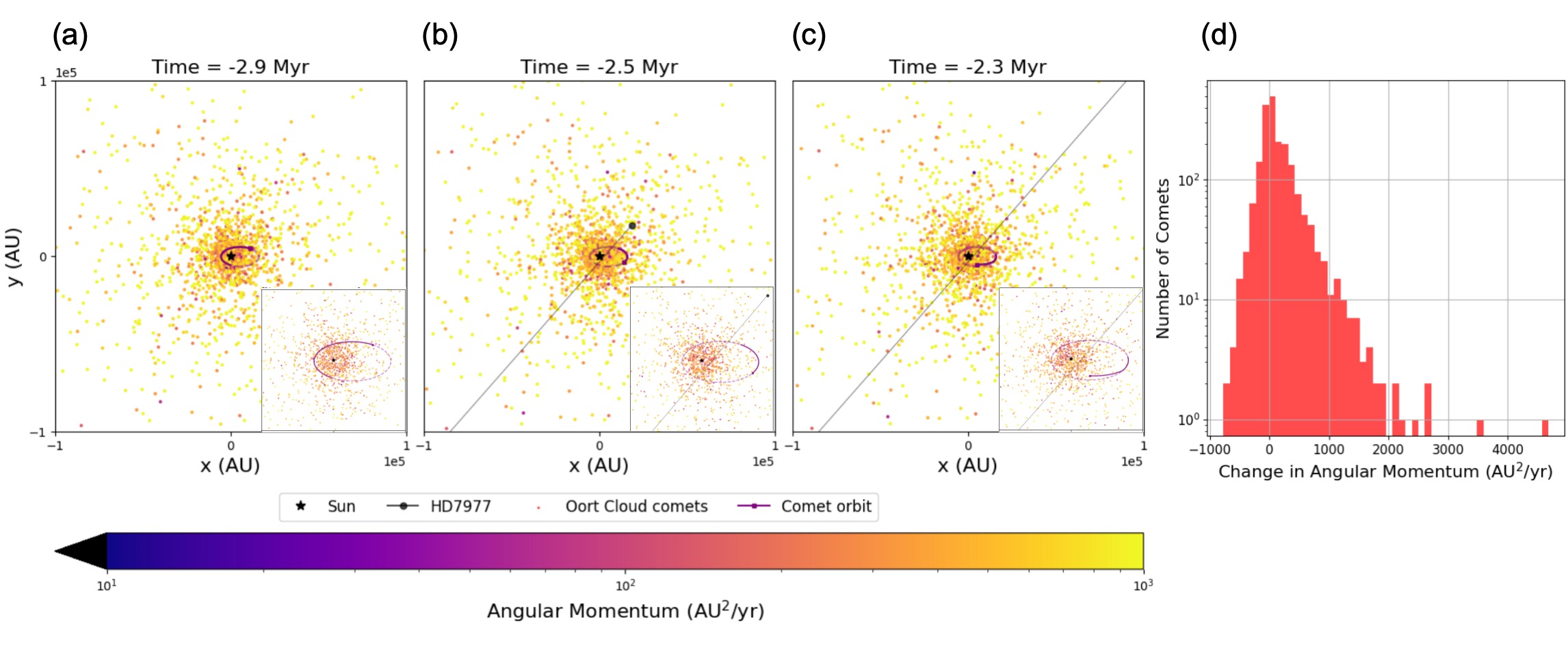}
\caption{Panel (a)-(c) is the snapshots of the REBOUND simulation. Colors represent the angular momentum value of each comet. Those with blue colors have a possibility of entering the inner solar system. The pentagram symbol denotes the Sun, the gray trajectory represents the flyby path of HD7977, and the purple trace illustrates the orbit of the selected comet. The insets inside each plot show the magnified orbit evolutionary of a selected representative comet. Panel (d) is the distribution of angular momentum change of Inner Oort cloud comets. }
\label{fig:simulation_result}
\end{figure}

Stellar flybys show significant dynamical perturbations to the Solar System's small body populations. 
In particular, HD7977's perturbation contributes to the dynamical repopulation of Oort Cloud comets. Figure \ref{fig:simulation_result} (a)-(c) shows the snapshots from the REBOUND simulation. The simulations were performed in three dimensions, with snapshots showing projections onto the x-y plane. However, due to the limited number of particles (1000 particles in IOC), these simulations cannot provide definitive conclusions; they are solely intended to depict general trends and present schematic illustrations for clarifying the subsequent theoretical framework. Following HD7977 flyby event, the orbital trajectories of selected representative comets exhibited a transformation, as evidenced by comparative analysis of their pre- and post-encounter configurations. The elliptical orbit depicted in Figure \ref{fig:simulation_result} (a) demonstrate an example comet's orbit prior to the significant perturbation from HD 7977, when its trajectory was still nearly Keplerian around the Sun. Following the gravitational interactions with the passing stellar body, the cometary path underwent orbital parameter modifications, manifesting as the reconfigured elliptical orbit with reduced perihelion distance as illustrated in Figure \ref{fig:simulation_result} (c). The observed orbital migration suggests energy transfer mechanisms between the cometary system and the transient gravitational potential of the flyby object, resulting in quantifiable changes to the affected bodies' Keplerian elements. Post-encounter analysis of the HD 7977 flyby event reveals redistribution of the comet angular momentum (as shown in Figure \ref{fig:simulation_result} (d)). There is an enhancement in cometary flux from the Oort Cloud, characterized by angular momentum depletion for some of the comets and consequent perihelion contraction across the perturbed population. Some of the comets with reduced perihelion distances enter the inner solar system. 

\begin{figure}[ht]
\centering
\includegraphics[width=\linewidth]{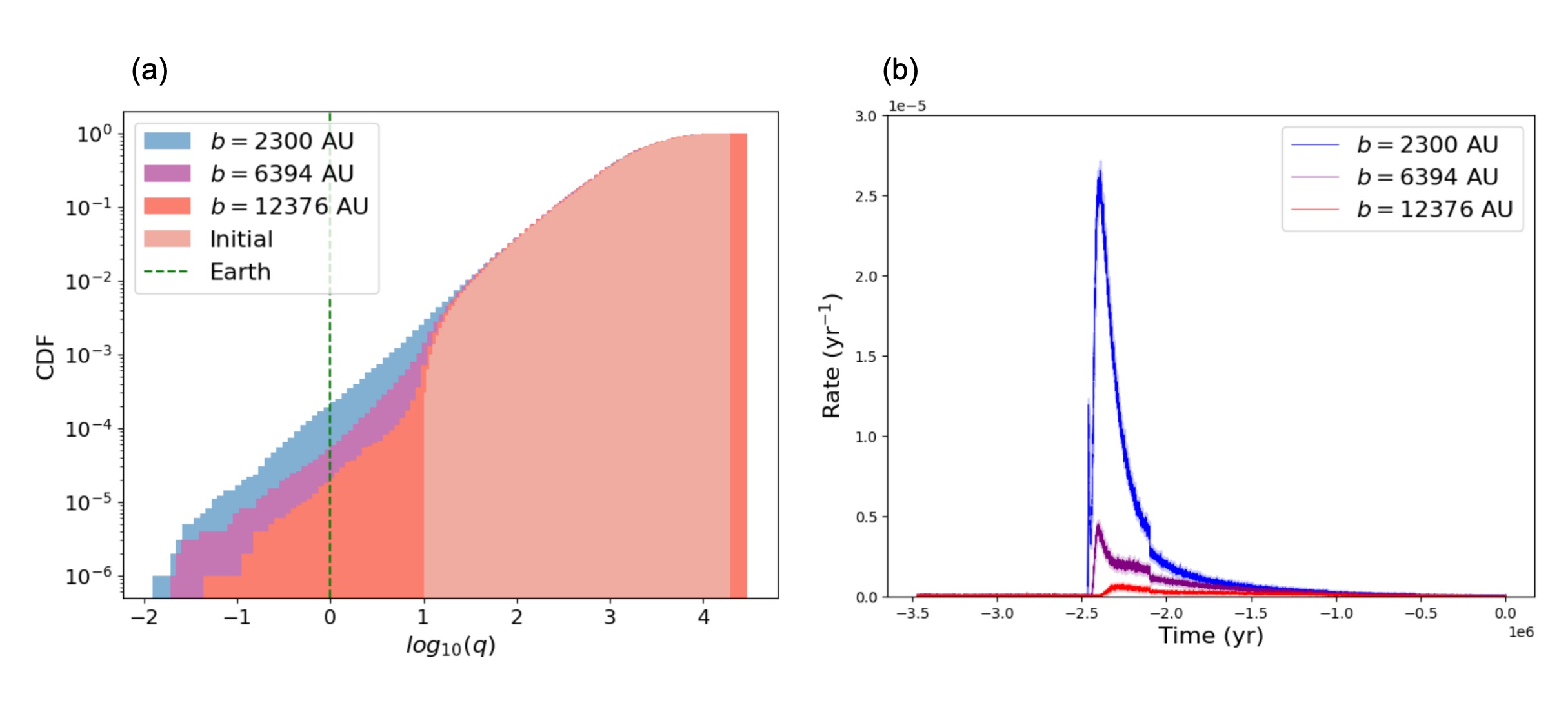}
\caption{(a) Distribution of $log_{10}(q)$ before and after the passage of HD7977 at different $b$, where $q$ represents the perihelion distance of comet orbits and $b$ represents the closest distance of HD7977 to the Sun. (b) Rate of comets with diameter $\geq 1~\text{km}$ colliding the Earth during the 3.5 Myr timescale. Year 0 represents the present (2025 A.D.). There was a peak of comet shower shortly after the passage of the HD7977 (-2.47 Myr). The transparent shaded region indicates the uncertainty in the results, which is relatively small. }
\label{fig:theory_discussion}
\end{figure}

To balance computational efficiency with analytical rigor, our work uses a combined approach of theoretical modeling with targeted numerical validation. For the theoretical calculations, the entire system was rotated to align the trajectory of HD 7977 with the y-axis to facilitate the application of the impulse approximation. This setup is illustrated in Figure \ref{fig:simulation_theory}(a) and Section `Theoretical calculation'. Our analysis assumes a steady-state configuration of the OOC over the 3.5 Myr timescale encompassed by this study, whereas the initial IOC before perturbation is characterized by an empty loss cone with perihelion distances of less than 10 AU \cite{weissman1985dynamical}, which is shown in Figure \ref{fig:theory_discussion} (a), labeled `Initial'. This setup reflects the understanding that galactic tides and distant stellar flybys can continuously supply new comets from the OOC to the inner solar system, while comets from the IOC are typically injected into the inner solar system during close stellar encounters \cite{hills1981comet} \cite{rickman2008injection}. Recent studies indicate that HD7977 might fly by with a perihelion distance of approximately $\sim$ 2300 AU from the Sun \cite{dybczynski2024hd}. Therefore, we focus our analysis on the 2300 AU close-encounter scenario, while alternative perihelion distances will be systematically addressed in the Discussion section. The perihelion distances of the perturbed IOC can be obtained by calculating the changes in angular momentum ($\vec{L}$) and orbit energy ($E$), 
\begin{equation}
\begin{aligned}
\vec{L}&=\vec{r} \times (\vec{v} + \Delta \vec{v}) \\
E &= -\frac{GM_\odot}{|\vec{r}|} + \frac{1}{2}|\vec{v} + \Delta \vec{v}|^2.
\label{eq:L_simple}
\end{aligned}
\end{equation}
By utilizing the post-perturbation angular momentum and orbital energy, the semi-major axis ($a$) and orbital eccentricity ($e$) of the modified trajectory can be calculated, thereby enabling the subsequent determination of the perturbed perihelion distance ($q$). The gravitational perturbation induced by the HD 7977 flyby significantly populated the loss cone of the inner Oort Cloud, as quantitatively demonstrated in Figure \ref{fig:theory_discussion} (a). 


If the loss cone is fully filled, the flux of comet shower could reach up to 37 times of its steady state, which is moderately higher than 20 times estimated by Duncan et al. (1987) \cite{duncan1987formation}. This discrepancy falls within an acceptable range and may result from the specific assumptions regarding the inner and outer Oort cloud (IOC/OOC) populations adopted in our model. Given Earth's orbital radius of 1 AU, our analysis focuses on comets penetrating the spherical shell defined by a 1 AU heliocentric radius. A comet can potentially approach within 1 AU of the Sun if its specific angular momentum satisfies:
\begin{equation}
\vec{L}=\vec{r} \times (\vec{v} + \Delta \vec{v}) \leq \left(\frac{GM_\odot}{1 \, \text{AU}}\right)^{1/2} \cdot (1 \, \text{AU})=\sqrt{GM_\odot 1 \, \text{AU}},
\label{eq:L_1}
\end{equation}
where the right-hand side represents the specific angular momentum for a circular orbit at 1 AU.

These comets exhibit non-negligible collisional potential with Earth, with planetesimal-scale impactors exceeding 1 km in diameter potentially posing globally meaningful consequences upon terrestrial impact. The Earth-impact rate for such events can be estimated as
\begin{equation}
\Gamma(t) = n(t)|v_{rel}|\sigma ,
\label{eq:rate}
\end{equation}
where $n$ is the comet number density inside the 1 AU sphere, $|v_{rel}|$ is the relative speed between the comet and the Earth and $\sigma$ is the cross-section of the Earth. 

To estimate the instantaneous density, $n$, within 1 AU, we solve for the orbits of perturbed, low-angular momentum comets. Application of Kepler's Third Law enables determination of ingress and egress timings for perturbed comets traversing the 1 AU heliocentric spherical shell, thereby allowing derivation of the instantaneous cometary number density ($n$) within this boundary. Here the calculation is based on the comets with diameter larger than 1 km. This density parameter is subsequently employed in Eq. \ref{eq:rate} to compute the Earth-impact rate ($\Gamma$), with resultant values exhibiting the temporal evolution shown in Figure \ref{fig:theory_discussion} (b). A notable enhancement in the Earth-impact rate ($\Gamma$) is found within the post-encounter regime, reaching levels above the steady-state baseline by over an order of magnitude. This transient enhancement appears as a peak in the temporal $\Gamma$-profile. Prior studies have suggested that cometary shower durations scale with the characteristic orbital period $P$ of perturbed cometary populations \cite{o2023pollution}. The 0.1 to 1 Myr timescale derived in our analysis is broadly consistent with this theoretical prediction, as the orbital period at 2300 AU is $\sim 10^5$~yr.


The net possibility of a collision with the Earth is calculated by integration over time, 
\begin{equation}
P=\int \Gamma(t)\text{d}t.
\label{eq:possibility}
\end{equation}
We adopt a characteristic interval of $T = 1$ Myr for the comet shower episode. When the cumulative impact probability attains $P = 1$ over this time, it implies one cometary collision with Earth within the 1 Myr interval. 

\begin{figure}[ht]
\centering
\includegraphics[width=\linewidth]{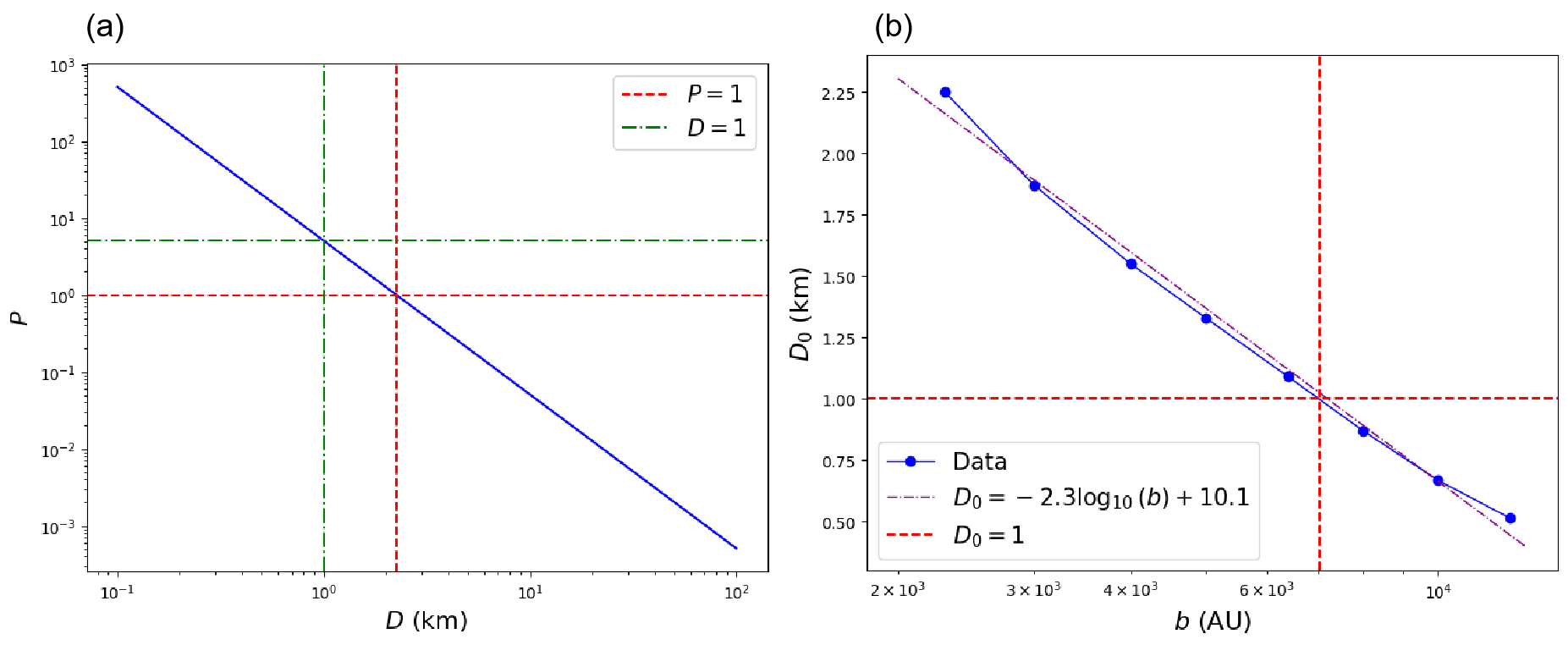}
\caption{(a) Collision probability ($P$) with Earth as a function of Oort Cloud comet diameter ($D$).
The critical diameter $D_0$ corresponding to a unit collision probability ($P=1$) is determined as 2.25 km. For $D=1$ km, the derived probability $P=5.08$ suggests an estimated $\sim 5$ potential Earth-colliding comets within the modeled parameter space. (b) Threshold diameter $D_0$ versus impact parameter $b$. The relationship exhibits a logarithmic dependence, yielding $b\approx7,000$ AU at $D_0=1$ km. }
\label{fig:theory_analysis}
\end{figure}

The size distribution of Oort Cloud comets roughly follows a power law $N \sim D^{-2}$, where $N$ is the comet population and $D$ is the diameter of the comet \cite{wajer2024oort}. Scaling the possibility according to the size distribution (Figure \ref{fig:theory_analysis} (a)) reveals that under the comet shower scenario, impactors with diameters up to 2.25 km are expected to collide with Earth within 1 Myr (cumulative probability $P = 1$), while $\sim$ 5 cometary impactors of kilometer-scale ($D \geq 1$ km) are estimated to collide over this interval. In contrast, the steady-state background flux suggests a maximum impactor diameter of 225 m, with a low terrestrial impact probability of $P \approx 0.05$ for 1 km-class bodies during the same timescale. Thus, if HD799 passed at a periapse distance of $\sim2300$~AU, an enhanced flux of IOC impactors is predicted by our modeling. The temporal coincidence of these impact events with the Pliocene-Pleistocene boundary indicates that geologic impact records from around this time could potentially be associated with a comet shower caused by the passage of HD7977.  

The perihelion distance of HD 7977's flyby is not well constrained by Gaia DR3, with dynamical reconstructions allowing for a close-approach range of 2300 to $\sim13,000$ AU \cite{dybczynski2024hd}. The resultant cometary bombardment intensity and associated Earth-impact probabilities depend on this flyby geometry. Figure \ref{fig:theory_discussion} illustrates three representative cases: a strong perturbation scenario ($b = 2,300$ AU), an intermediate case ($b = 6,394$ AU), and a weak perturbation scenario ($b = 12,376$ AU).

As shown in Figure \ref{fig:theory_discussion} (a), all three encounter geometries result in partial replenishment of the loss cone, though the replenishment efficiency decreases as the closest approach distance $b$ increases. For the intermediate case ($b = 6,394$ AU), the critical threshold – defined as the maximum comet diameter with cumulative probability $P = 1$ within 1 Myr – decreases to 1.09 km, with an estimated total number of kilometer-scale impactors ($D \geq 1$ km) of approximately 1–2. In the distant encounter scenario ($b = 12,376$ AU), this critical diameter further diminishes to 514 m, while the collision probability for 1 km-class comets reduces to $P \approx 0.3$. The threshold diameter $D_0$ exhibits a strong dependence on HD 7977's closest approach distance $b$, as shown in Figure \ref{fig:theory_analysis} (b). For $b \leq 7,000$ AU, the cumulative impact probability exceeds unity for 1 km impactors over the 1 Myr interval. 

Another uncertainty to consider is that our parametrization of the relative IOC and OOC population densities would affect the flyby-induced comet flux.\cite{hills1981comet} \cite{wajer2024oort} \cite{kaib2008formation}. We have assumed a relatively low IOC density. If actual populations are higher than these estimates, the corresponding Earth-impact probabilities would be increased. Conversely, an overestimation of the population densities would lead to a proportional downward revision of the impact risks, with our final impact-flux results depending approximately linearly on the assumed IOC to OOC relative density.




 

\section*{Discussion}


Collectively, our dynamical models are consistent with the hypothesis that HD 7977's late Pliocene flyby  initiated a comet shower episode temporally coincident with the Pliocene-Pleistocene transition. Model results indicate that gravitational perturbations from this flyby would have substantially enhanced Earth's impactor flux, with the potential for sub-kilometer to kilometer-scale collisions over 1 Myr. 

We emphasize that our results on the intensity of the comet shower and the subsequent Earth-impact probability are model-dependent, particularly on the assumed initial structure of the inner Oort cloud (IOC). According to Figure 7 of Vokrouhlicky (2019) \cite{vokrouhlicky2019origin}, the inclination distribution for the IOC appears more concentrated near the ecliptic plane compared to the broader distribution (0–180°) characteristic of the OOC. This configuration could potentially arise because the distribution of eccentricities may reach thermal equilibrium during dynamical evolution, while the `memory' effect of inclination may persist longer. Within the semi-major axis range of the IOC, external perturbations such as Galactic tides and past stellar encounters may be comparatively less efficient at exciting inclination than at modifying perihelion distance, though the precise mechanisms require further investigation. 
Our fiducial model assumes a disk-like IOC. However, due to the inherent uncertainties in the Oort Cloud model, we also tested a spherical model as a variant \cite{kaib2008formation, 1997Icar..129..106F}. Since the flyby trajectory of HD 7977 is inclined relative to the ecliptic plane, the fiducial disk-like structure does not appear to substantially enhance the probability of a reduction in the perihelion distance $q$ compared to the spherical case.
As a result, the results from both models yield generally consistent outcomes.

The asymmetric nature of our model (studying only comets initial bound to the Sun) raises an important consideration regarding the possibility that HD 7977 itself may possess an extrasolar Oort cloud. Although HD 7977 has lower metallicity than the Sun, stars with comparable metallicity are known to host planetary systems \cite{2007ApJ...665.1407C}, and likely also cometary reservoirs. Given the flyby geometry, the inner Solar System would have traversed any cometary cloud associated with the intruding star, with the Sun's gravity potentially focusing these impactors. 
To provide a rough estimate, we consider the relative contribution from HD 7977's hypothetical Oort cloud. The impact flux from an extrasolar Oort cloud can be estimated as:
\begin{equation}
\Phi_{\text{exo}} \propto n_{\text{exo}} \cdot v_{\text{enc}} \cdot \sigma_{\text{focus}} \cdot P_{\rm strip}(M_{\odot},M_{\text{exo}})
\end{equation}
where $n_{\text{exo}}$ is the number density of comets in HD 7977's Oort cloud, $v_{\text{enc}}$ is the encounter velocity, $\sigma_{\text{focus}}$ is the gravitational focusing cross-section, and $P_{\rm strip}(M_{\odot},M_{\text{exo}})$ is the stripping probability. 
A comparison between the induced flux from solar and extrasolar sources can to somewhat show the contribution of HD7977's comets. We provide here a rough estimate of the comet flux. Given that the binding energy of Oort cloud particles is very small, we approximate it as negligible compared to the energy change induced by stellar perturbations. The stripping probability scales as $P_{\rm strip}(M_{\odot}, M_{\text{exo}}) \propto (M_{\odot}/M_{\text{exo}})^2$. Assuming the encounter velocity $v_{\text{enc}}$ and the focusing cross-section $\sigma_{\text{focus}}$ are comparable for this system, and that the number density of exo-Oort clouds $n_{\text{exo}}$ is approximately equal to the local density $n_{\odot}$, the exo-comet flux can be estimated as $\Phi_{\text{exo}} \approx \Phi_{\odot} \times (M_{\odot}/M_{\text{exo}})^2$. Based on our calculation, this yields $\Phi_{\text{exo}} \sim 4~\text{Myr}^{-1}$ for impactors larger than 1 km when the impact parameter $b = 2300$ AU. 

Notably if a comet shower did occur --  peaking shortly after HD7977's passage (Figure 2b) -- the resultant impact winter mechanisms coincide temporally with the Pleistocene shift. The Pliocene-Pleistocene transition is generally regarded as a gradual transition over Myr, rather than an impulsive episode \cite{clark2024global}. In the context of a collision-induced model, this gradual nature would be more consistent with an enhanced flux of impactors predicted by a comet shower (Fig \ref{fig:theory_discussion}) than with a single, catastrophic collision. 

The lunar records provide complementary perspectives on temporal variations in impact flux due to the absence of erosion. Mazrouei et al. (2019) used Diviner thermophysical data to calibrate large lunar crater ages and identified a statistically significant increase in crater production beginning at ~290 Ma, implying a contemporaneous rise in the flux of large impactors to Earth \cite{mazrouei_earth_2019}. Yang et al. (2020) expanded the lunar crater catalog through automated recognition and relative age assignment, thereby enhancing the statistical power to detect short-term variations in impact flux across different size ranges \cite{yang_lunar_2020}. However, we note that most studies of lunar craters focus on timescales of gigayears, with relatively scarce research on the several-megayear range. Future geological studies that identify a peak in the lunar cratering rate around 3 Myr ago could provide supporting evidence for a comet shower induced by HD 7977's flyby. 

We also surveyed terrestrial impact craters with nominal ages between 10 and 0 Ma to identify potential candidates linked to the proposed increase in impactor flux. There are a series of impact craters discovered that have the possibility of forming between 3 Ma to 1 Ma. It is critical to note the significant uncertainties associated with crater ages. Many impacts go unrecorded because they occur in oceans or are masked by erosion, sedimentation, and vegetation, while strict validation criteria often prevent the recognition of airbursts and oceanic impacts\cite{masse2007missing}. Both recorded and unrecorded impacts may have something to do with the comet shower caused by HD 7977. According to the Earth Impact Database maintained at the Planetary and Space Science Center at the University of New Brunswick, other famous craters in the same period are El'gygytgyn in Russia (4 Ma to 3 Ma, 18 km in diameter) \cite{gurov2007gygytgyn}, Zhamanshin in Kazakhstan (1 Ma to 0.8 Ma, 14 km in diameter) \cite{schulz2020zhamanshin}, Bosumtwi in Ghana (1.07 Ma, 10.5 km in diameter) \cite{jones1981lake}, Karla in Russia (6 Ma to 4 Ma, 10 km in diameter) \cite{quesnel2022karla}, and Bigach in Kazakhstan (8 Ma to 2 Ma, 8 km in diameter)\cite{komatsu2019bigach}. We can deduce the diameters of the related impactors to be $\sim$ 1 km to 2 km, which are enough to cause global catastrophes. Asteroid-caused Eltanin impact is thought to occur in the South Pacific Ocean around the Pliocene-Pleistocene boundary \cite{gersonde1997geological}, which is already suggested as a contributor to the climate changes at the transition \cite{goff2012eltanin}. The impactor in Eltanin is thought to have a diameter of 1 to 4 km \cite{gersonde1997geological}. While the prevailing view attributes the Pliocene–Pleistocene transition primarily to asteroidal impacts, the findings presented here suggest a plausible alternative scenario involving cometary bombardment. Further geological investigations will be needed to evaluate the potential role of such a mechanism. 

Under typical impact flux conditions, the catastrophic impactor threshold for asteroids within a 1 Myr interval ranges between 1-2 km in diameter \cite{michel2015asteroids}. Our calculations suggest that during the HD7977-induced comet shower episode, the terrestrial impact probabilities for cometary bodies approach those of those of asteroidal impactors. Notably, geological evidence does not preclude a cometary origin for the aforementioned impact events. This ambiguity, combined with our dynamical modeling results, lends support to the possibility that some of these impacts could potentially have originated from cometary sources. In Section ``Method--Comets and Asteroids'' we have also analyzed the power-law size distributions of comets and asteroids, finding that comets likely dominate at smaller sizes.

The detectability of a comet shower associated with HD 7977 in the geological record depends on its intensity exceeding the background rate of asteroid impacts. Distinguishing between asteroidal and cometary impact craters remains challenging. Furthermore, some comets may disintegrate in the atmosphere, leaving no crater. More recently, Fernández (2025) modeled asteroid-belt mass loss and showed that extrapolated fluxes are highly sensitive to present-day loss parameters, offering a framework for testing whether observed flux increases can be explained by asteroidal depletion alone or require additional cometary sources \cite{fernandez_depletion_2025}. According to Mazrouei et al. (2019), the lunar impact flux capable of producing craters larger than 20 km in diameter over the past 85 Myr is $2.5 \times 10^{-15}$ km$^{-2}$yr$^{-1}$. Converted to Earth's cross-sectional area, this corresponds to approximately 1.28 impacts per Myr, which can be regarded as the background asteroid impact rate\cite{mazrouei_earth_2019}. In our study, for an impact parameter $b = 2300$ AU, the resulting impact flux reaches about 4–5 per Myr—a value higher than the asteroid background and thus may be detectable above background levels. Our models indicate that a comet shower could have affected Earth within approximately $\pm 0.5$ Myr of 2 Ma. Future geological studies that confirm such a signal would provide support for the potential role of the HD 7977 flyby in the Pliocene–Pleistocene transition.

In conclusion, the HD 7977-induced comet shower may have influenced Earth's impactor flux during the Pliocene-Pleistocene transition, and could have been a factor in the environmental perturbations associated with this geological boundary interval.




\section*{Methods}

\subsection*{Basic settings}

The dynamical calculations were performed using a heliocentric unit system with solar mass ($M_\odot$), astronomical unit (AU), and year (yr) as base units, yielding a gravitational constant of 
$G=39.4769 \text{AU}^3/(M_\odot\text{yr}^2)$. The stellar perturber HD 7977 was modeled with a mass of $M_p=1.07M_{\odot}$ , exhibiting a heliocentric relative velocity of $v = 26.45$~km/s. The velocity components in the Cartesian reference frame were specified as $v_x=11.67$ km/s, $v_y=13.27$ km/s, and $v_z=19.68$ km/s. The minimum heliocentric distance was adopted as 2300 AU, consistent with recent astrometric reconstructions \cite{dybczynski2024hd}.

For comets in the OOC, the orbital elements were initialized as follows \cite{vokrouhlicky2019origin}:
\begin{equation*}
\begin{aligned}
&\bullet \text{Eccentricity } e \text{ follows a thermalized distribution } (f(e) = 2e); \\
&\bullet \text{Semi-major axis } a \text{ was sampled from a log-uniform distribution spanning } 20000 \text{ to } 100000~\text{AU}; \\
&\bullet \text{Inclination } i \text{ was isotropically distributed } (\cos i \sim \mathcal{U}(-1,1)); \\
&\bullet \text{Angular elements } M, \omega, \Omega \sim \mathcal{U}(0, 2\pi), \\
&\quad \text{where } M \text{ denotes mean anomaly, } \omega \text{ the argument of perihelion, and } \Omega \text{ the longitude of ascending node.}
\end{aligned}
\end{equation*}

For the IOC comets, the orbital parameters were initialized as follows \cite{vokrouhlicky2019origin}:
\begin{equation*}
\begin{aligned}
&\bullet \text{Eccentricity } e \text{ follows a thermal distribution } (f(e) = 2e); \\
&\bullet \text{Semi-major axis } a \text{ was sampled from a log-uniform distribution spanning } 2000 \text{ to } 20000~\text{AU}; \\
&\bullet \text{A planetary scattering truncation was imposed at perihelion distance } q = 10~\text{AU}; \\
&\bullet \text{Inclination distribution bifurcated by semi-major axis:} \\
&\quad - \text{For } a \in (2000, 5000)~\text{AU: } i \sim \mathcal{N}(\pi/6, \pi/18) \\
&\quad - \text{For } a \in (5000, 20000)~\text{AU: } \cos i \sim \mathcal{U}(-1, 1)  \\
&\quad \text{where the notation }\mathcal{N}(\mu, \sigma) \text{ denotes a normal (Gaussian)  probability distribution characterized by mean }\mu \\
&\quad\text{ and standard deviation } \sigma \text{, and }\mathcal{U}(a, b) \text{ means a uniform probability distribution on }(a, b); \\
&\bullet \text{Angular elements } M, \omega, \Omega \sim \mathcal{U}(0, 2\pi), \\
&\quad \text{where } M \text{ denotes mean anomaly, } \omega \text{ the argument of perihelion, and } \Omega \text{ the longitude of ascending node.}
\end{aligned}
\end{equation*}

Our simulation employed a total of $10^6$ test particles to represent the inner Oort cloud in Figure \ref{fig:theory_discussion}(a). Following the stellar encounter, approximately 0.02\% of the simulated particles evolved onto Earth-crossing orbits ($q < 1$ AU) for the $b=2,300$ AU case, while for the 6,394 AU and 12,376 AU cases, the fractions are reduced to 0.005\% and 0.001\%. 

The statistical uncertainty associated with our finite sample size can be estimated using binomial statistics. For an impact probability p derived from our simulation (e.g., the fraction of particles that become Earth-crossing), the standard error is approximately $\sqrt{p(1-p)/N}$. For $b=2,300$ AU, $p = 0.02\%$  and N = $10^6$, this translates to an uncertainty of $\pm 0.0015\%$. For the 6,394 AU and 12,376 AU cases, the uncertainties of $p$ are $\pm 0.00007\%$ and $\pm 0.00003\%$. 

For Figure \ref{fig:theory_discussion}(b), in order to reduce statistical uncertainties in counting the number of particles entering within 1 AU at each time step, a sample of $10^{10}$ IOC particles was employed in the calculation, with the results scaled to the realistic value of $5\times 10^{12}$. The transparent shaded region represents the uncertainty in the results, and it can be observed that statistical uncertainty is small relative to the collision probability. 


\subsection*{REBOUND simulation to verify the theory}

The parameters for the REBOUND simulation shown in Figure \ref{fig:simulation_result} were configured as mentioned above, with specific settings established for both the IOC and OOC, as well as parameters corresponding to HD7977. Due to computational time constraints, the number of simulated particles could not be set at a high value, prompting the adoption of theoretical calculations to enhance computational efficiency. In these theoretical computations, we assumed the OOC to remain in a steady state while focusing exclusively on the IOC due to its susceptibility to perturbations caused by HD7977's flyby. The impulse approximation was applied in the theoretical framework, and we first validated its feasibility through comparative analysis between simulation results and impulse approximation predictions.

\begin{figure}[ht]
\centering
\includegraphics[width=\linewidth]{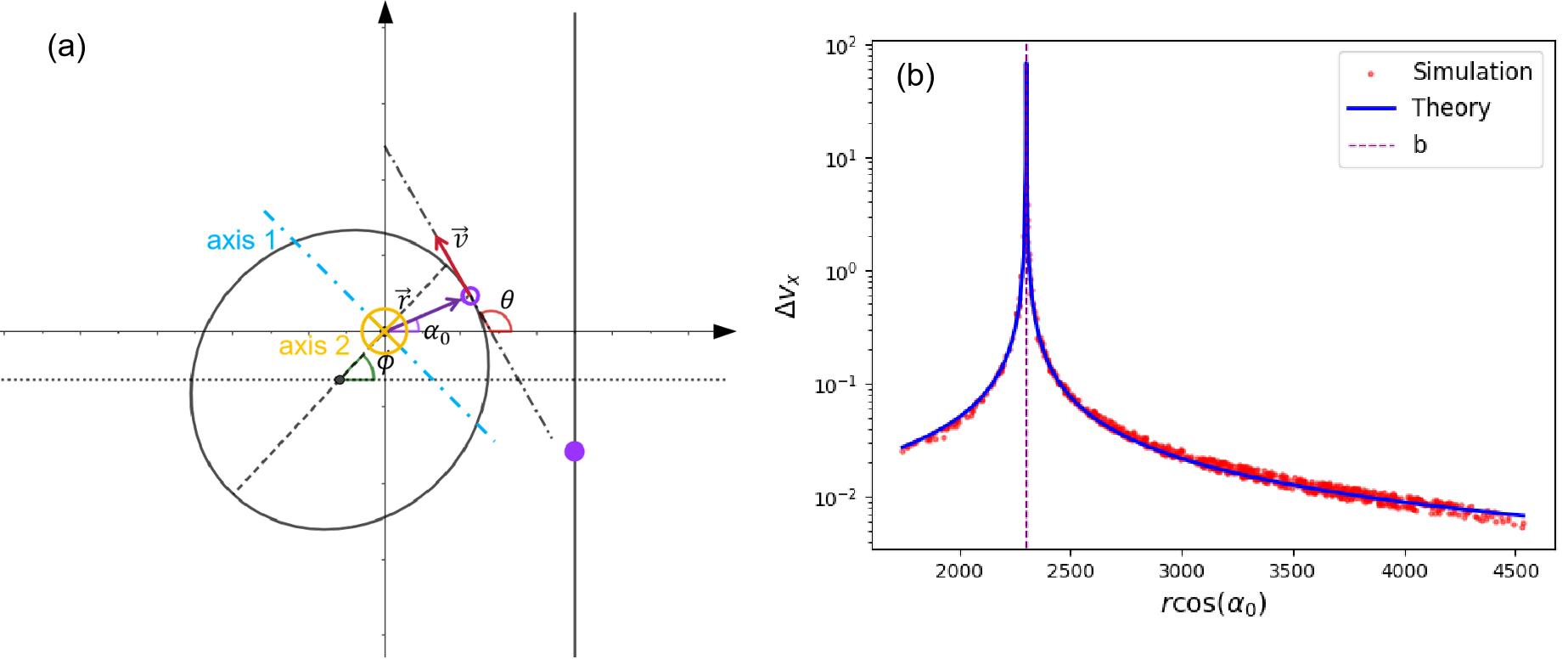}
\caption{(a) Schematic of orbital parameters, where $\vec{r}$ is the position vector of a comet, $\vec{v}$ is the velocity vector, $\alpha_0$ is the angle between $\vec{r}$ and the x-axis, $\theta$ is the angle between $\vec{v}$ and the x-axis, $\phi$ is the orientaion of the ellipse orbit and axis 1 and axis 2 are two rotation axes. (b) Comparison between impulse approximation theory and simulation, using $|\Delta v_x|$ as a metric.}
\label{fig:simulation_theory}
\end{figure}

Complementary REBOUND simulations were implemented  to validate core theoretical assumptions -- impulse approximation. From impulse approximation, $\Delta v$ is calculated as: 
\begin{equation}
\begin{aligned}
\dot{\vec{v}} &= GM_p\frac{(b-r\cos\alpha_0, y_{p0}+v_pt-r\sin\alpha_0)}{((b-r\cos\alpha_0)^2+(y_{p0}+v_pt-r\sin\alpha_0)^2)^{3/2}} \\
\Delta\vec{v} &=  \int_{-\infty}^{+\infty}  \dot{\vec{v}}\, dt\\
&=(\frac{2GM_p}{v_p}\frac{1}{b-r\cos\alpha_0},0),
\label{eq:delta_v}
\end{aligned}
\end{equation}
where $G$ is gravitational constant, $M_p$ is the mass of HD7977 in solar mass, $v_p$ is the relative velocity to the Sun of HD7977, $y_{p0}$ is the initial $y$ coordinate of HD7977 and $b$ is the closest approaching distance of HD7977 to the Sun. The parameters associated with the comet are shown in Figure \ref{fig:simulation_theory} (a). 

The REBOUND simulation framework was configured with three celestial bodies: the Sun, HD7977, and a randomly selected Oort Cloud particle. Since none of these particles were in close proximity to the central Sun during the perturbation phase, the `whfast' integrator was employed to optimize computational efficiency. To simplify the dynamical configuration, the Oort Cloud particle was assigned a circular orbit, while HD7977's trajectory was prescribed perpendicular to the $x$-axis. This resulted in HD7977's initial 2D Cartesian coordinates being $(b, -y_{p0})$ with a velocity component $v_{py}=v=26.45km/s$. The Oort Cloud particle's $x$-coordinate was randomly sampled near $b$, with its orbital true anomaly $f$ uniformly distributed between 0 and $2\pi$. Following the calculation of orbital radius through the relation $r=x/\cos(f)$, particles exceeding $10^4$ AU or falling below 2000 AU in orbital radius were systematically excluded from the analysis, according to the radius range of the IOC comets.

To determine the orbital parameters of the Oort Cloud particle at closest approach, we computationally solved 
\begin{equation}
\begin{aligned}
r\sin(f+\omega t)&=-y_{p0}+v_{py}t \\
f'&=f+\omega t \\
x' &= r\cos(f')
\label{eq:encounter_time}
\end{aligned}
\end{equation}
to derive the true anomaly $f'$ and corresponding $x'$-coordinate. Particles exhibiting non-physical solutions ($t<0$), indicating absence of actual stellar encounter, were systematically eliminated through this analytical filtering process.

As demonstrated in Figure \ref{fig:simulation_theory} (b), using $|\Delta v_x|$ as a metric, the close agreement between REBOUND simulation and theoretical predictions confirms the applicability of impulse approximation within our defined parameter space, particularly for encounters whose timescale is much shorter than the orbital period of the perturbed body, allowing the gravitational interaction to be treated as an instantaneous momentum transfer. This methodological synergy ensures both computational tractability and physical consistency across all analytical domains.

\subsection*{Theoretical calculation}

Within the theoretical framework, our analysis focused exclusively on the IOC component, with parameter configurations for both HD7977 and IOC maintained as specified in "Basic settings" to ensure methodological consistency. 

For an IOC comet, as shown in Figure \ref{fig:simulation_theory} (a), the true anomaly satisfies $f = \alpha_0 - \phi$, where $\alpha_0$ denotes the initial angular displacement. The REBOUND function M\_to\_f was employed to numerically resolve the $\alpha_0$ distribution. 

The geometrical representation of an ellipse in Cartesian coordinates is characterized by its semi-major axis $a$ and semi-minor axis $b$, formally expressed as ${x^2}/{a^2} + {y^2}/{b^2} = 1$. The equation of the tangent line at an arbitrary point $(x_0, y_0)$ on the ellipse is mathematically derived as $y = -{b^2x_0}/{a^2y_0}~(x - x_0) + y_0$. When analyzing elliptical orbital dynamics, where the position vector of the celestial body is denoted $\vec{r} = (x, y)$, the polar angle $\theta$ relative to the semi-major axis is determined through: 
\begin{equation}
\tan(\theta - \phi) = -\frac{b^2}{a^2} \frac{c + r \cos(\alpha_0 - \phi)}{r \sin(\alpha_0 - \phi)}.
\label{eq:tangent_line}
\end{equation}
This formulation incorporates $\alpha_0$, representing the polar coordinate angle, and $\phi$, defining the rotational orientation of the elliptical trajectory about its principal axes. 

The angular displacement $\theta$ of the celestial body is derived through the relation $\theta = \phi + \arctan\left(-\frac{b^2}{a^2}\frac{c + r \cos(\alpha_0 - \phi)}{r \sin(\alpha_0 - \phi)}\right)$, where $\phi$ denotes the ellipse's orientation angle. In the polar coordinate framework, the radial position $r$ of the Oort Cloud particle follows the standard elliptical orbital solution $r = \frac{a(1 - e^2)}{1 + e \cos(\alpha_0 - \phi)}$, where $a$ specifies the semi-major axis and $e$ represents the orbital eccentricity. The instantaneous orbital velocity $v$ as a function of radial distance $r$ is governed by the energy conservation principle, 
\begin{equation}
\begin{aligned}
-\frac{GM_\odot}{2a}&=-\frac{GM_\odot}{r}+\frac{1}{2}v^2 \\
v &= \sqrt{GM_\odot\left(\frac{2}{r} - \frac{1}{a}\right)}.
\label{eq:v}
\end{aligned}
\end{equation}
The Cartesian position and velocity vectors are mathematically defined as $\vec{r} = (r \cos \alpha_0, r \sin \alpha_0)$ and $\vec{v} = (v \cos \theta, v \sin \theta)$. The orbital plane was subsequently transformed through two successive rotational operations, as diagrammed in Figure \ref{fig:simulation_theory} (a): 

1. Primary rotation: About Axis 1 (through the foci, parallel to the semi-minor axis) by inclination angle $i$.

2. Secondary rotation: About Axis 2 (orthogonal to the transformed plane post-first rotation, also passing through foci) by argument of periapsis $\omega$. 

3. Final rotation: Through this coordinate system transformation, the position vectors ($\vec{r}$) and velocity vectors ($\vec{v}$) of Oort Cloud particles were co-rotated with HD7977's velocity orientation to align the stellar perturber's velocity vector with the (0,1,0) axis. This strategic frame alignment optimized the calculation of velocity perturbation ($\Delta v$) derivation while preserving the kinematic relationships between the Oort Cloud population and HD7977.

The orientation of Axis 2 was dynamically determined through the cross product $\vec{r} \times \vec{v}$ computed after the primary rotation. Defining unit vectors $\hat{e}_1$ (Axis 1) and $\hat{e}_2$ (Axis 2), the composite rotational transformation is governed by the matrix:
\begin{equation}
\begin{aligned}
\mathbf{R}(\hat{e}_1, i) &= \cos(i) \mathbf{I}_3 + (1 - \cos(i)) \hat{e}_1 \hat{e}_1^T + \sin(i) [\hat{e}_1]_\times \\
\mathbf{R}(\hat{e}_2, \omega) &= \cos(\omega) \mathbf{I}_3 + (1 - \cos(\omega)) \hat{e}_1 \hat{e}_1^T + \sin(\omega) [\hat{e}_1]_\times.
\label{eq:matrix}
\end{aligned}
\end{equation}
Here, $[\hat{e}_1]_\times$ represents the skew-symmetric matrix (or cross-product matrix) corresponding to the vector $\hat{e}_1$. This matrix is used to represent the cross product as a matrix multiplication operation. For a 3D vector $\hat{e}_1 = \begin{bmatrix} x \\ y \\ z \end{bmatrix}$, the skew-symmetric matrix $[\hat{e}_1]_\times$ is defined as:
$$
[\hat{e}_1]_\times = 
\begin{bmatrix}
0 & -z & y \\
z & 0 & -x \\
-y & x & 0
\end{bmatrix}.
$$
Given HD7977's three-dimensional velocity vector $\vec{v} = (11.67,\,13.27,\,19.68)$ km/s in the heliocentric frame, we implemented rotational transformations to ensure the proper orientation of the flyby passage to the Oort Cloud. The rotational transformation matrix from HD7977's velocity vector $\vec{v}_p$ to the reference direction $(0,1,0)$ was computed. Given initial unit vectors $\hat{v}_1 = \vec{v}_p/\|\vec{v}_p\|$ and target vector $\hat{v}_2 = (0,1,0)$, the rotation axis $\hat{e}_3$ was determined by $\hat{e}_3 = \frac{\hat{v}_1 \times \hat{v}_2}{\|\hat{v}_1 \times \hat{v}_2\|}$. The rotation angle $\beta$ satisfies $\cos\beta = \hat{v}_1 \cdot \hat{v}_2$, $\sin\beta= \|\hat{v}_1 \times \hat{v}_2\|$. Special cases were analytically handled. For general cases, the rotation matrix was derived by:
\begin{equation}
\mathbf{R}(\hat{e}_3, \beta) = \cos(\beta) \mathbf{I}_3 + (1 - \cos(\beta)) \hat{e}_3 \hat{e}_3^T + \sin(\beta) [\hat{e}_3]_\times.
\label{eq:dv_rotation}
\end{equation}
Then,
\begin{equation}
\begin{aligned}
\vec{r'} &= \mathbf{R}(\hat{e}_3, \beta)\mathbf{R}(\hat{e}_2, \omega)\mathbf{R}(\hat{e}_1, i)\vec{r} \\
\vec{v'} &= \mathbf{R}(\hat{e}_3, \beta)\mathbf{R}(\hat{e}_2, \omega)\mathbf{R}(\hat{e}_1, i)\vec{v}.
\label{eq:rv}
\end{aligned}
\end{equation}
The velocity perturbation ($\Delta v$) applied to the orbiting body of an Oort Cloud particle initially at $(x_0, y_0, z_0)$ after the perturbation of HD7977 can be calculated as:
\begin{equation}
\begin{aligned}
\dot{\vec{v}} &= GM_p\frac{(b-x_0, y_{p0}+v_pt-y_0,z_0)}{((b-x_0)^2+(y_{p0}+v_pt-y_0)^2+z_0^2)^{3/2}} \\
\Delta\vec{v} &=  \int_{-\infty}^{+\infty}  \dot{\vec{v}}\, dt\\
&=(\frac{2GM_p(b-x_0)}{((b-x_0)^2+z_0^2)v_p},0,\frac{-2GM_pz_0}{((b-x_0)^2+z_0^2)v_p}).
\label{eq:delta_v_3D}
\end{aligned}
\end{equation}
Here, $M_p$ represents the perturber' mass, $v_p$ denotes the heliocentric velocity of HD7977, and $b$ specifies the closest flyby distance. Also, as for the close flyby, HD7977 can perturb solar motion in the Galaxy and change the orbits of all bodies of the Solar System at different scales \cite{dybczynski2024hd}. So $\Delta v$ in Eq. \ref{eq:delta_v_3D} was subsequently corrected by subtracting the solar motion component $\Delta v_\text{sun}$ through coordinate system alignment. $\Delta v' = \Delta v - \Delta v_\text{sun}$, where $\Delta v_\text{sun}$ represents the heliocentric velocity adjustment calculated using identical method as Eq. \ref{eq:delta_v_3D}. 



We computed the specific orbital energy $E$ for each Oort Cloud particle after the perturbation. This energy parameter enables direct determination of the semi-major axis $a$ through the fundamental relationship expressed as follows: 
\begin{equation}
\begin{aligned}
\vec{r}_{\text{after}} &= \vec{r} \\
\vec{v}_{\text{after}} &= \vec{v} + \Delta \vec{v} \\
\vec{L}_{\text{after}} &= \vec{r}_{\text{after}} \times \vec{v}_{\text{after}} \\
E_{\text{after}} &= -\frac{GM_\odot}{|\vec{r}_{\text{after}}|} + \frac{1}{2}|\vec{v}_{\text{after}}|^2 \\
a_{\text{after}} &= -\frac{GM_\odot}{2E_{\text{after}}}.
\label{eq:after}
\end{aligned}
\end{equation}
From unit angular momentum and semi-major axis, we can calculate the eccentricity and perihelion distance $q$:
\begin{equation}
\begin{aligned}
q_{\text{after}} &= a_{\text{after}}(1-e_{\text{after}}) \\
-\frac{GM_\odot}{2a_{\text{after}}}&=-\frac{GM_\odot}{q_{\text{after}}}+\frac{1}{2}v_{\text{peri}}^2 \\
|\vec{L}_{\text{after}}| &= v_{\text{peri}}q_{\text{after}}.
\label{eq:q}
\end{aligned}
\end{equation}
So,
\begin{equation}
\begin{aligned}
e_{\text{after}} &= \sqrt{1 - \frac{|\vec{L}_{\text{after}}|^2}{GM_\odot a_{\text{after}}}} \\
q_{\text{after}} &= \frac{|\vec{L}_{\text{after}}|^2}{(1+e_{\text{after}})GM_\odot}.
\label{eq:q_}
\end{aligned}
\end{equation}

\subsection*{Falling time calculation and collision possibility}


The temporal evolution of perturbed IOC comets with post-encounter perihelia $q_{\text{after}} \leq 1$ AU requires careful dynamical analysis. The transition timescale $\Delta t$ from perturbation to actual reaching at 1 AU is fundamentally determined by the mean anomaly $M_{\text{after}}$, which in turn depends on the true anomaly $f_{\text{after}}$ through Keplerian orbital mechanics. The post-perturbation true anomaly is derived from the orbital geometry relation $r_{\text{after}} = \frac{a_{\text{after}}(1 - e_{\text{after}}^2)}{1 + e_{\text{after}} \cos(f_{\text{after}})}$. This formulation enables direct computation of $\cos(f_{\text{after}})$: 

\begin{equation}
f_{\text{after}} = \begin{cases} \arccos(\cos(f_{\text{after}})) & \text{if } \vec{r}_{\text{after}} \cdot \vec{v}_{\text{after}} > 0, \\ 
2\pi - \arccos(\cos(f_{\text{after}})) & \text{otherwise}. 
\end{cases}
\label{eq:f_after}
\end{equation}

Mean anomaly $M$ is a key parameter describing the fraction of an orbit's period that has elapsed since periapsis. The calculation is dependent on the orbital eccentricity $e$ and the true anomaly $f$.

For $e$ and $f$, we identify the orbit type (elliptical or hyperbolic) and invoke the appropriate sub-function to calculate $M$: 
\begin{equation}
M(e, f) =
\begin{cases} 
2 \arctan\left(\sqrt{\frac{1 - e}{1 + e}} \tan\left(\frac{f}{2}\right)\right) - \frac{e \sqrt{1 - e^2} \sin(f)}{1 + e \cos(f)} & \text{if } e \in (0, 1), \\[10pt]
\frac{e \sqrt{e^2 - 1} \sin(f)}{1 + e \cos(f)} - \ln\left(\frac{\sqrt{e + 1} + \sqrt{e - 1} \tan\left(\frac{f}{2}\right)}{\sqrt{e + 1} - \sqrt{e - 1} \tan\left(\frac{f}{2}\right)}\right) & \text{if } e > 1.
\end{cases}
\label{eq:M_after}
\end{equation}
If $M < 0$ for elliptical orbits, the result is adjusted by adding $2\pi$ to ensure $M$ remains within the range $[0, 2\pi]$.

For cometary particles exhibiting post-perturbation perihelion distances $q \leq 1$ AU, we determined the critical mean anomalies $M_{\text{in}}$ (1 AU crossing ingress) and $M_{\text{out}}$ (1 AU crossing egress) through the same computational framework established in Eq. \ref{eq:M_after}. The characteristic angular velocity $\omega$ was derived from $\omega = \sqrt{\frac{GM}{|a|^3}}$. The transition timescales were then computed through distinct formulations based on orbital regime: For elliptical orbits, $t_{\text{ell, in}} = -t_p + \frac{|M_{\text{in}} - M|}{\omega}$. For hyperbolic orbits, 
\[
t_{\text{hyp, in}} = 
\begin{cases}
-t_p + \frac{|M_{\text{in}} - M|}{\omega} & \text{if } M < 0 \\
\infty & \text{if } M \geq 0
\end{cases}. 
\]
where $\infty$ denotes the absence of perihelion passage. The egress timescale $t_{\text{out}}$ was computed through analogous methodology.

We systematically recorded the 1 AU crossing durations for all dynamically viable cometary trajectories. To quantify the temporal distribution of cometary flux, we implemented a statistical sampling approach by uniformly distributing $10^7$ temporal evaluation points across the interval from -3.47 Myr to present epoch. The computational analysis employed segment tree algorithm to efficiently determine the instantaneous cometary number density within the 1 AU sphere at each temporal sample point. The resulting number density profile was subsequently transformed into an impact rate distribution through application of the scaling relation defined in Eq. \ref{eq:rate}. This processed dataset yielded the theoretical cometary flux evolution presented in Figure \ref{fig:theory_discussion} (b), providing quantitative insight into the temporal variation of inner Solar System cometary activity.

To quantify the impact probability of Oort Cloud comets with Earth, we established a population model based on current understanding of Oort Cloud demographics, incorporating an estimated $10^{12}$ comets with diameters $D \geq 1$ km in the outer Oort Cloud and $5 \times 10^{12}$ in the inner Oort Cloud. The size distribution follows a power-law relation $N(D) \propto D^{-2}$, where $D$ represents the cometary diameter and $N(D)$ the differential size distribution. The terrestrial impact probability $P_{\text{impact}}$ was computed through the relation expressed in Eq. \ref{eq:possibility}. 


The relative velocity $v_{\text{rel}}$ in Eq. \ref{eq:possibility} requires careful computation, particularly given the characteristic infall velocity of 8.88 AU/yr for Oort Cloud particles reaching 1 AU, which is comparable to the Earth orbiting velocity. To determine the average relative velocity between Earth and an incoming Oort Cloud particle, we modeled their interaction probabilistically, assuming isotropic velocity vector distributions. Letting $v_{\text{Earth}}$ and $v_{\text{Oort}}$ represent the respective velocity magnitudes, with $\theta \in [0, \pi]$ denoting the uniformly distributed angle between their velocity vectors, the instantaneous relative velocity follows $v_{\text{rel}} = \sqrt{v_{\text{Earth}}^2 + v_{\text{Oort}}^2 - 2v_{\text{Earth}}v_{\text{Oort}}\cos\theta}$. The expectation value of $v_{\text{rel}}$ was computed by integrating over the uniform angular distribution $f(\theta) = \frac{1}{2\pi}$, yielding the average relative velocity:  
\begin{equation}
\langle v_{\text{rel}} \rangle = \frac{1}{\pi} \int_{0}^{\pi} \sqrt{v_{\text{Earth}}^2 + v_{\text{Oort}}^2 - 2v_{\text{Earth}}v_{\text{Oort}}\cos\theta} \, d\theta.
\label{eq:v_average}
\end{equation}
This framework enables  analysis of cometary impact probabilities with Earth. Through analysis on dynamics of Oort Cloud, we establish a probabilistic model for quantifying terrestrial impact risks from long-period comet during comet shower cause by a stellar flyby.  

\subsection*{Comets and Asteroids}

Furthermore, the analysis of size-frequency distributions reveals distinct scaling laws for different small body populations: Recent studies suggest that Oort Cloud comets follow a size-frequency distribution of $N_{\text{comet}}(> D) \propto D^{-1.4}$ to $D^{-1.6}$ for $D>1$ km populations \cite{2019Icar..333..252B}\cite{2017AJ....154...53B}, while some earlier work proposed a steeper slope of $D^{-2}$. Near-Earth asteroids typically exhibit a distribution of $N_{\text{asteroid}}(> D) \propto D^{-1.3}$\cite{2021AJ....162..280M}. More recent research indicates that asteroids with $D > 100$ m may follow a shallower distribution of $N_{\text{asteroid}}(> D) \propto D^{-0.97}$ \cite{burdanov2025jwst}. Since our work yields comparable impact fluxes for 1 km-scale objects from the Oort Cloud and the asteroid belt, the relatively steeper slope of the cometary size distribution suggests a potentially higher relative abundance of smaller impactors ($0.2< D < 1$ km) from the Oort Cloud compared to the asteroid belt. Although the steeper SFD for comets indicates a greater proportion of smaller objects, our analysis of potentially significant environmental effects focuses primarily on larger impactors capable of producing global-scale consequences.

\bibliography{sample}



\section*{Data availability}
The codes and datasets analyzed during the current study are archived online\cite{cao_2025_15600997}.

\section*{Acknowledgements}


The authors wish to express their gratitude to Dr. Julio A. Fernandez, Dr. Jakub Rozehnal, and an anonymous reviewer for their thought-provoking comments that greatly facilitated the refinement of this paper. Z.C. thanks Yang Liu for his support during the research, Prof. Shude Mao and Dr. Yukun Huang for valuable discussions, and the Tsien Excellence Engineering Program (TEEP). 

\section*{Author contributions statement}
Z.C. carried out the analysis and prepared the manuscript. A.L. conceptualized the study, and assisted with interpretation. M.M. assisted with the analysis and interpretation.

\section*{Funding}
Z.C. was supported by the Tsien Excellence in Engineering Program at Tsinghua University. M.M. gratefully acknowledges support from a Clay Postdoctoral Fellowship at the Center for Astrophysics, Harvard \& Smithsonian. 





\end{document}